\documentclass[aps,prc,superscriptaddress,twoside,twocolumn,nofootinbib,showpacs]{revtex4-1}

\usepackage{amsmath,amssymb}
\usepackage{mathtools}
\usepackage{bbold}
\usepackage[usenames,dvipsnames]{xcolor}
\usepackage{braket}
\usepackage{graphicx,bm}
\usepackage{slashed}
\usepackage{tensor}

\usepackage{multirow,physics}

\allowdisplaybreaks

\begin{document}

\title{\boldmath Nuclear energy density functional and the nuclear $\alpha$ decay}

\author{Yeunhwan \surname{Lim} }
\email{ylim@tamu.edu}
\affiliation{Cyclotron Institute and Department of Physics and Astronomy, 
Texas A\&M University, College Station, Texas 77843, USA}

\author{Yongseok \surname{Oh}}
\email{yohphy@knu.ac.kr}
\affiliation{Department of Physics, Kyungpook National University, Daegu 41566, Korea}
\affiliation{Asia Pacific Center for Theoretical Physics, Pohang, Gyeongbuk 37673, Korea}

%\date{\today}

\begin{abstract}
The nuclear $\alpha$ decay of heavy nuclei is investigated based on the nuclear energy density 
functional, which leads to the $\alpha$ potential inside the parent nucleus in terms
of the proton and neutron density profiles of the daughter nucleus. 
We use the Skyrme force model, Gogny force model, and relativistic mean field model to get 
the nucleon density profiles inside heavy nuclei.
Once the nucleon density profiles are determined, the parameters of the nuclear $\alpha$ potential
are fitted to the observed $\alpha$ decay half-lives of heavy nuclei.
This approach is then applied to predict unknown $\alpha$ decay half-lives of heavy nuclei.
To estimate the $Q$ values of unobserved $\alpha$ decays, we make use of the 
liquid droplet model.
\end{abstract}


\maketitle

%%%%% Section 1

\section{Introduction}
The synthesis of unknown heavy nuclei has been spotlighted for last decades with the 
development of new facilities for rare isotope accelerators~\cite{Moller16,GS16,GGW16}.
In particular, the structure of neutron-rich heavy nuclei is expected to shed light on
our understanding of nuclear structure in isospin asymmetric nuclear matter and it will
give insight on the structure of neutron stars and the process of nuclear synthesis during 
the evolution of stars~\cite{GLM07}.
Therefore, it can be a test ground for various issues of nuclear physics such as nuclear 
density functional, strong nuclear interactions, various decay processes, r-p process, etc, 
which makes it one of the most exciting topics in low energy nuclear physics~\cite{OST17}.
The formation of such heavy nuclei is identified through their decay processes such as the
$\alpha$ decay, $\beta$ decay, and spontaneous fission~\cite{VS66b}. 
The competition between these decay processes is reflected in branching ratios, and,
in fact, the heavy nuclei with the atomic number $Z>105$ were found to rarely survive for 
a few minutes~\cite{AWWK12, WAWK12}.

The study on the nuclear $\alpha$ decay process has a very long history, as it is one of 
the major decay processes of nuclei~\cite{Mang64,VS66b}.
In particular, the formation of a new heavy nuclide would be mostly identified through its $\alpha$ 
decay chains~\cite{Oganessian07,MMKH12,KYDA14}.
Modern approaches for theoretical understanding of the nuclear $\alpha$ decay 
are based on effective nuclear interactions such as the square well potential 
model~\cite{BMP91,BMP92}, $\cosh$ potential model~\cite{BMP92b},
unified fission model~\cite{DZGWP10}, double-folding model~\cite{RSB05,SCB07,RSB08}, 
and so on.

The most important factor in the $\alpha$ decay process of heavy nuclei is the accurate 
information on the $Q$ value for the decay process, which reflects the structure of heavy 
nuclei through binding energy.
The importance of the $Q$ value in the $\alpha$ decay lifetime can easily be found in the
Geiger-Nuttall law~\cite{GN11} and its improved version of Viola and Seaborg~\cite{VS66b}.%
\footnote{
For example, in the case of the alpha decay of $\nuclide[212]{Po} \to \nuclide[208]{Pb} + \alpha$, 
a difference of 0.1 MeV in the $Q$ value of the reaction, where $Q_{\rm expt.} \approx 8.95$~MeV, 
results in about a factor of 1.7 difference in the calculated half-life of \nuclide[212]{Po}.}

The next most sensitive factor in the determination of the $\alpha$ decay width is the nucleon 
distribution inside the daughter nucleus, which determines the $\alpha$ potential. 
Since the $\alpha$-decay is basically a quantum tunneling effect, the exact positions of the 
classical turning points and the profile of the barrier, i.e., its height and width, are essential
parts for the estimation of the $\alpha$ decay lifetime. 
Therefore, the information on the nuclear potential felt by the $\alpha$ cluster inside the parent 
nucleus is important to estimate the $\alpha$ decay width.
Furthermore, the Coulomb potential is responsible for the repulsive potential barrier together with the
angular momentum barrier, so the potential shape due to the proton distribution in the daughter 
nucleus has a nontrivial role in the $\alpha$ decay process.
The purpose of the present work is to go beyond a simple model approach for the $\alpha$ 
potential by developing a more realistic $\alpha$ potential based on nucleon density profiles
for estimating $\alpha$ decay half-lives.

In the present work, we calculate the $\alpha$ decay half-lives of heavy nuclei within the 
Wentzel-Kramers-Brillouin (WKB) approximation by calculating the nuclear potential felt by
the $\alpha$ cluster using phenomenological nuclear force models.
The nuclear potential form for the $\alpha$ cluster is obtained from the Skyrme-type interaction 
as prescribed in Ref.~\cite{SLHO15}, which requires the proton and neutron distribution functions 
as inputs.
We then use the Skyrme SLy4 model~\cite{CBHMS98} and Gogny D1S model~\cite{BGG91} as 
non-relativistic models and the relativistic mean-field DD-ME2 model~\cite{LNVR05} as well. 
For the $Q$ values of the $\alpha$ decay processes, we use the experimental data whenever
available, and, if not, we make use of the liquid droplet model (LDM) elucidated in Ref.~\cite{SPLE04}.

This paper is organized as follows. 
In Sec.~\ref{sec:qval}, we review the LDM to calculate the binding energy to be used when
the experimental $Q$ value is not known. 
The Coulomb diffusion and exchange terms are included as well as the pairing and
shell corrections, which gives a better fitting to existing data. 
In the shell corrections, we use the last magic number as a free variable to minimize
the root-mean-square deviation of total binding energy. 
We also check the $Q$ values using the phenomenological formula as a function of isospin 
asymmetry $I$, with $I=(N-Z)/A$, as in Ref.~\cite{DZGWP10}.
Section~\ref{sec:models} briefly explains nuclear models to find the density profiles of nucleons
inside nuclei, and we construct the effective nuclear potential for the $\alpha$ cluster.
The parameters of the effective potential for each model of nucleon density distribution
are determined.
Our results are presented in Sec.~\ref{sec:results} and compared to experimental data. 
The predictions on the unobserved $\alpha$ decays of heavy nuclei are given as well.
We summarize and conclude in Sec.~\ref{sec:conclusion}.

\section{\boldmath Nuclear models}
\label{sec:qval}

The $Q$ value plays an important role to determine the lifetime of the $\alpha$ decay  
as it determines the assaulting frequency of the $\alpha$ particle for a given potential well. 
It also sets the penetration width for quantum tunneling.
In the estimation of $\alpha$ decay lifetimes, we use the empirical $Q$ values, if available.
However, for unobserved decay processes, we have to resort to model predictions on the 
binding energy.
In this Section, we review the LDM that will be used in the present work.

\subsection{Liquid Droplet Model}

To estimate the unknown binding energies of heavy nuclei, we make use of the LDM with 
some modifications as prescribed in Refs.~\cite{MS69,SPLE04}.
In general, heavy nuclei are neutron-rich and the neutron skin is likely to exist on the surface. 
For example, the neutron skin thickness of \nuclide[208]{Pb} was investigated with
the electric dipole response~\cite{TPVF11},
the parity radius experiment (PREX)~\cite{PREX12, HAJR12},
and, more recently, through coherent $\pi^0$ photoproduction~\cite{CB14}.
All numerical calculations using Skyrme-Hartree-Fock, Gogny, and relativistic mean
field models show the out-layer of neutrons in the neutron-rich heavy nuclei.
Thus, it is natural to include the neutron skin effects in LDM.
The binding energy in the LDM for a nucleus of ($Z, A$) is given as~\cite{SPLE04}
\begin{eqnarray}\label{eq:ldm}
E & = & f_{B}^{} \left( A - N_{s} \right) + 4\pi R^{2} \sigma(\mu_{n}) + \mu_{n}N_{s} + E_{\text{Coul}} 
\nonumber \\ && \mbox{}
+ E_{\text{pair}} + E_{\text{shell}},
\end{eqnarray}
where $f_B^{}$ is the binding energy per baryon of infinite nuclear matter,
$N_s$ is the number of neutrons in the neutron skin on the surface, 
$R$ is the radius of the nucleus,
$\sigma (\mu_n)$ is the surface tension as a function of neutron chemical potential $\mu_n$.
$E_\text{Coul}$ is the Coulomb energy, $E_\text{pair}$ is the pairing energy, and 
$E_\text{shell}$ includes the shell corrections.

In this model, $f_B$ is a phenomenological energy function, which reads
\begin{equation}
f_B^{}  = - B + S_v(1-2x)^2 + \frac{K}{18}(1-u)^2,
\end{equation}
where $B$ is the binding energy per nucleon, $S_v$ is the nuclear symmetry energy,
and $K$ is the nuclear incompressibility of symmetric nuclear matter at nuclear saturation 
density $\rho_0^{}$. 
Here, $x$ and $u$ are defined as
\begin{equation}
x = \frac{Z}{A-N_s} , \quad u =\frac{\rho}{\rho_0^{}}.
\end{equation}
The surface tension is a function of $x$, and we find that the simple expansion of 
$\sigma(x) = \sigma_0^{} - \sigma_\delta^{} (1-2x)^2$
is not a good approximation for highly neutron-rich nuclei.
Therefore, we use the form suggested in Refs.~\cite{RPL83, Lim12},
which reads
\begin{equation}
\sigma(x) = \sigma_0^{} \frac{2\cdot 2^\alpha + q}{x^{-\alpha} + q + (1-x)^{-\alpha}}\,.
\end{equation}
The parameters $\sigma_0^{}$, $\alpha$, and $q$ will be determined later.

The Coulomb energy contribution to the total mass is obtained from 
the classical Coulomb interaction, the Coulomb diffusion term, and
the exchange term.
It is then written as
\begin{equation}
E_\text{Coul} = \frac{3Z^2e^2}{5R}
-\frac{\pi^2 Z^2 e^2 d^2}{2R^3}
- \frac{3Z^{4/3}e^2}{4R} \left(\frac{3}{2\pi}\right)^{2/3}\,,
\end{equation}
where $d$ ($= 0.55~\text{fm})$ is the surface diffuseness parameter~\cite{SPLE04} and
$R$ is the average radius of the nucleus.
The general expression for the pairing energy in LDM reads
\begin{equation}
E_\text{pair} = (-1)^N \frac{\Delta_N}{\sqrt{A}}
+ (-1)^Z\frac{\Delta_P}{\sqrt{A}}\,,
\end{equation}
where the pairing energies for protons and neutrons are treated separately,
since neutron-rich nuclei would have higher single particle energy of the last-filled 
neutron than the one for protons.

For the shell contribution to the total binding energy, we follow the prescription of 
Duflo and Zuker~\cite{DZ94, DV09}, which writes the shell correction as
\begin{equation}
E_\text{shell}
= a_1^{} S_2 + a_2^{} (S_2)^2 + a_3^{} S_3 + a_{np}^{} S_{np}\,,
\end{equation}
where $a_1^{}$, $a_2^{}$, $a_3^{}$, and $a_{np}^{}$ are parameters to be determined,
and
\begin{equation}
\begin{aligned}
S_{2} & = \frac{n_v \bar{n}_v}{D_n} + \frac{p_v \bar{p}_v}{D_p} \,, \\
S_{3} & = \frac{n_v \bar{n}_v (n_v - \bar{n}_v)}{D_n} 
         + \frac{p_v \bar{p}_v (p_v - \bar{p}_v)}{D_p} \,, \\ 
S_{np} & = \frac{n_v \bar{n}_v p_v \bar{p}_v}{D_nD_p} \,.
\end{aligned}
\end{equation}
Here, $n_v$ and $p_v$ are the valence numbers of neutrons and protons, respectively, 
i.e., the minimal difference for neutron and proton numbers from the magic numbers, 
2, 8, 20, 28, 50, 82, 126, and 184 (or 168). 
For example, for \nuclide[56]{Fe}, we obtain $n_v = \abs{30 - 28} = 2$ and $p_v = \abs{26 - 28} = 2$.
$D_n$ ($D_p$) is the degeneracy number, i.e, the interval of the magic numbers adjacent to
the neutron (proton) number. 
For instance, in the case of \nuclide[56]{Fe}, the nearest two magic
numbers for $N=30$ are $28$ and $50$, which then leads to $D_{N=30} = 50 - 28 = 22$.
Finally, $\bar{n}_v$ and $\bar{p}_v$ are the complementary valence numbers for neutrons and
protons, respectively, and their explicit forms are
\begin{equation} 
\bar{n}_v \equiv D_n - n_v, \quad \bar{p}_v \equiv D_p - p_v.
\end{equation}
Again, for \nuclide[56]{Fe}, we have $\bar{n}_v(30) = 22 - 2 = 20$.

\begin{table}[t]
\begin{tabular}{cccl}
\hline
\hline 
                & ~~Case I~~             &  ~~Case II~~ &    ~Unit \\
\hline
$B$             &  16.125                     &  16.370       &    ~MeV \\
$\rho_0^{}$     &  0.155                   &  0.155        &   ~fm$^{-3}$ \\
$\sigma_0^{}$   &  1.256                 &  1.300        &   ~MeV\,fm$^{-2}$ \\
$\alpha$        &  4.0                    &  3.7          &   \\
$q$             &  60.00                  &  25.48        &   \\
$S_v$           &  31.818                   &  32.471       &   ~MeV\\
$K$             &  250.00               &  226.389      &   ~MeV\\
$\Delta_n$      &  5.458                    &  6.232        &   ~MeV\\
$\Delta_p$      &  5.807                     &  11.760       &   ~MeV\\
$a_1^{}$        &  1.265                   & $-0.143$     &   ~MeV\\
$a_2^{}$        & $-8.601\times 10^{-3}$     &  $9.307\times 10^{-3}$ &  ~MeV\\
$a_3^{}$        &  $-4.007\times 10^{-3}$    &  $2.216\times 10^{-3}$ &  ~MeV\\
$a_{np}^{}$     &  $-9.663\times 10^{-2}$   &   $-4.231\times 10^{-2}$   &   ~MeV\\
$M(8)$          &  184                      &  168          & \\
RMSD            &  1.144                        &  0.218      &   ~MeV\\
\hline
\end{tabular}
\caption{The parameters of LDM.
The values of case I are obtained by the least $\chi^2$ fitting to the observed
binding energies for 2336 nuclei.
The parameters in case II are found by fitting to the experimental $Q$ values for the nuclei 
with $Z \ge 100$, where we have totally 100 data points.
$M(8)$ is the 8th magic number in each case.
RMSD in the last row denotes the root-mean-square deviation.
The RMSD in cases I is for binding energies, whereas that in case II is for $Q$ values.
}
\label{table1}
\end{table}

In the present work, we will work with two parameter sets as given in Table~\ref{table1}.
The parameters of case I are obtained by fitting to the experimentally known binding energies of 
2336 nuclei. 
Therefore, this corresponds to a global fitting.
On the other hand, since we are considering $\alpha$ decays of neutron-rich heavy nuclei, 
it may be useful to focus on heavy nuclei for that purpose.
Thus the second parameter set is found by using the measured $Q$ values of heavy nuclei 
with $Z \ge 100$.
We use 100 data points for finding the parameters set of case II.
Note that $M(8)$ in Table~\ref{table1} is the 8th magic number in the LDM parameterization with
each parameter set.

Once the masses of nuclei are evaluated by Eq.~\eqref{eq:ldm}, we can calculate the 
$Q$ value for $\alpha$ decay through~\cite{MRDT06}
\begin{eqnarray}\label{eq:qval}
Q &=&  \Delta M(Z,A) - \Delta M(Z-2, A-4) - \Delta M_{\alpha}
\nonumber \\ && \mbox{}
+ 10^{-6}\, k\left[ Z^\beta - (Z-2)^\beta \right] ,
\end{eqnarray}
where $\Delta M_\alpha = 2.4249$~MeV.
The values for $k$ and $\beta$ are ($k=8.7$~MeV, $\beta = 2.517 $)
for nuclei of $Z \ge 60$, and ($k=13.6$~MeV, $\beta =2.408 $)
for nuclei of $Z < 60$.

\subsection{\boldmath Local formula for $Q_\alpha$}

Considering heavy nuclei with $Z \ge 90$ and $N \ge 140$, Dong et al.~\cite{DR08,DZGWP10} 
developed a local mass formula for nuclei with large $N$ and $Z$ values.
Using the Taylor expansion, it leads to the expression of the local $Q$ value including shell effects as
\begin{eqnarray}\label{eq:qform}
Q & = & a \frac{Z}{A^{4/3}} \left(3A-Z \right) + b\left( \frac{N-Z}{A} \right)^2 
\nonumber \\ && \mbox{}
+ c \left[ \frac{\abs{N-152}}{N} - \frac{|N-154|}{N-2}\right]
\nonumber \\ && \mbox{}
+ d \left[ \frac{|Z-110|}{Z} - \frac{|Z-112|}{Z-2}\right] + e ,
\end{eqnarray}
where $a$, $b$, $c$, $d$, and $e$ are parameters to be fitted.
Note that the pairing effects are neglected since the semi-classical formula 
gives almost the same contribution to the total binding energy for both
parent and daughter nuclei and it does not cause a change in the $Q$ value.
Since our goal is to compute the half-lives of super heavy nuclei through $\alpha$ decay 
processes, we obtain the parameters in Eq.~\eqref{eq:qform} with the measured $Q$ values
for nuclei with $Z \ge 100$.
The resulting parameters are shown in Table~\ref{tab2}.

\begin{table}[t]
\begin{tabular}{cccccc}
\hline\hline
$a$         & $b$    & $c$   & $d$   & $e$  & RMSD \\
\hline
$0.90753$   & $-97.84028$  & $16.15924$  & $-18.95722$  & $-26.16600$ & $0.255$ \\
\hline
\end{tabular}
\caption{The best fit parameters of Eq.~\eqref{eq:qform}. 
All parameters have a unit of MeV.}
\label{tab2}
\end{table}

\begin{figure*}[t]
\includegraphics[scale=0.65]{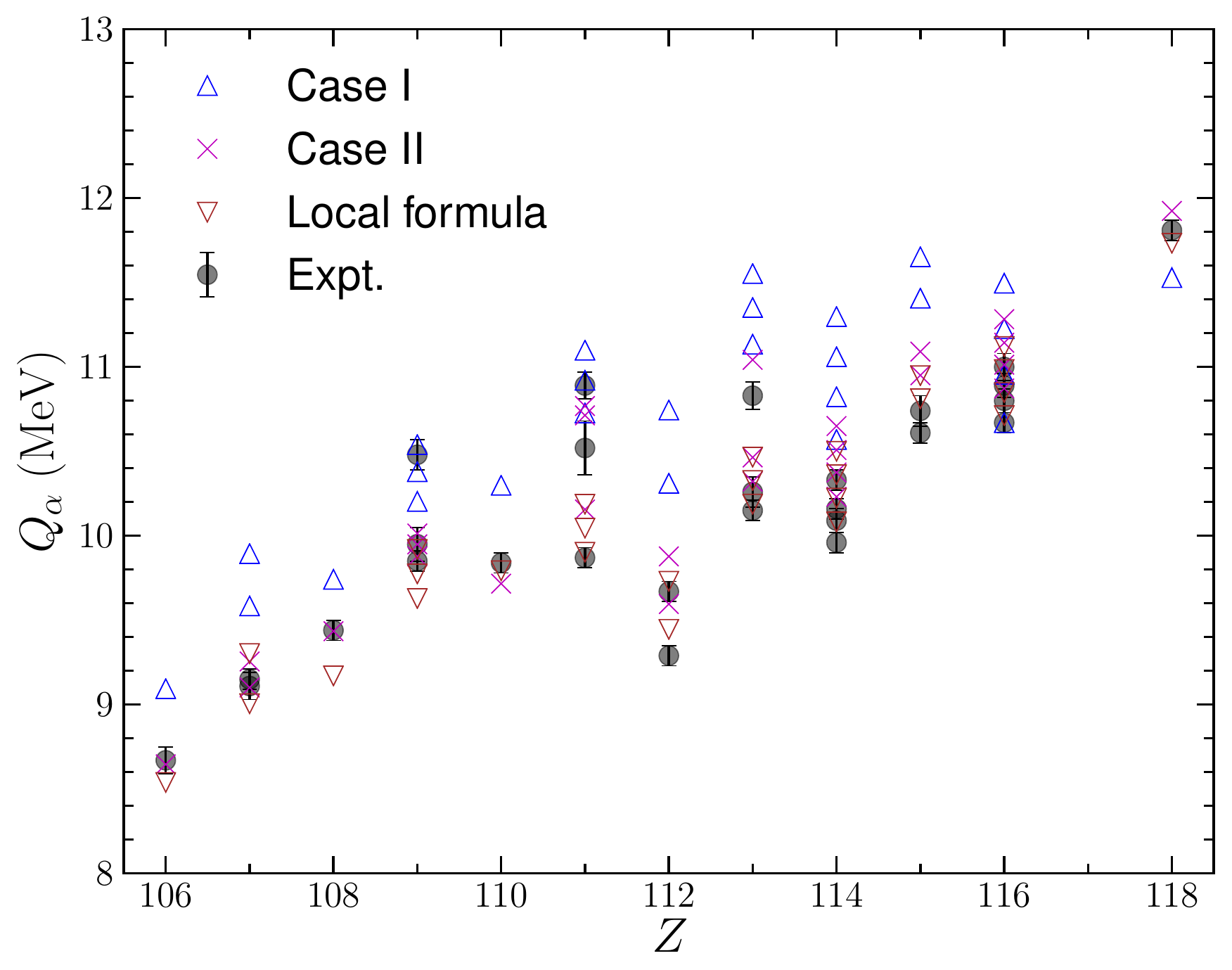}
\caption{$Q$ values for $\alpha$ decays of nuclei between $Z = 106$ and $Z = 118$. 
The numerical values can be found in Table~\ref{tb:heavy}.
A small horizontal offset is used for better visibility for a given value of $Z$.}
\label{fig:qval}
\end{figure*}

Figure~\ref{fig:qval} shows the $Q$ values obtained from the LDM with 
Eq.~\eqref{eq:qval} and those from the local formula of Eq.~\eqref{eq:qform}. 
It is found that the case II and the local formula give more reliable results 
than case I on the measured $Q$ values.

\section{\boldmath Potential for the $\alpha$ cluster}\label{sec:models}

In the $\alpha$ cluster model, the nuclear $\alpha$ decay is described as a quantum tunneling effect.
Once the energy, i.e., the $Q$ value, of the reaction is determined, the next
step is to find the potential for the $\alpha$ cluster inside the parent nucleus.
In this Section, we discuss how we use phenomenological models for constructing
the potential for the $\alpha$ cluster.

\subsection{\boldmath Potential form}

In the $\alpha$ cluster model, the $\alpha$ particle is already formed in the parent nucleus and
it penetrates the potential barrier to cause the $\alpha$ decay process.
Therefore, the estimation of lifetimes requires the information on the potential of the 
$\alpha$ cluster created by the core nucleus, i.e., the daughter nucleus after decay.

The $\alpha$ cluster potential can be decomposed as
\begin{equation}
V = V_N + V_C + V_L,
\label{eq:V_alpha}
\end{equation}
where $V_N$ is the nuclear potential for the $\alpha$ cluster,
$V_C$ is the Coulomb potential provided by the protons of the core nucleus,
and $V_L $ is the centrifugal potential arising from the relative orbital angular momentum
between the $\alpha$ particle and the core nucleus.
In principle, the nuclear potential of the $\alpha$ particle would be computed if the interactions 
between nucleons inside a nucleus is completely known.
However, it is certainly beyond the scope of the present work, and we invoke the Skyrme force 
model to get the form of $V_N$.
Then, as described in Ref.~\cite{SLHO15}, $V_N$ takes the form of
\begin{eqnarray}
V_N & = &  \alpha \rho + \beta(\rho_n^{5/3} + \rho_p^{5/3}) 
+ \gamma \rho^{\epsilon}(\rho^2 + 2\rho_n^{} \rho_p^{}) 
\nonumber \\ && \mbox{}
+ \delta \frac{1}{r}\frac{d \rho}{dr}  + \eta\frac{d^2 \rho}{dr^2} \,,
\label{eq:V_N}
\end{eqnarray}
where $\rho = \rho_n^{} + \rho_p^{}$ with $\rho_n^{}$ ($\rho_p^{}$) being the density distribution of
neutrons (protons).
This model contains 6 parameters, namely, $\alpha$, $\beta$, $\gamma$, $\delta$,
$\eta$, and $\epsilon$.
These parameters will be determined by fitting to the empirical data for $\alpha$ decay 
half-lives of heavy nuclei and will be discussed in the next subsection.
Furthermore, the nuclear potential in Eq.~\eqref{eq:V_N} is controlled by the density
distribution of nucleons, which should be provided by microscopic models for nuclear structure.

Once the nucleon distribution is known, the Coulomb potential term $V_C$ can be calculated
through
\begin{equation}
V_C = 8\pi e^2 \left[ \frac{1}{r} \int_0^r \rho_p^{}(r^\prime) r^{\prime2} dr^\prime 
+ \int_r^\infty \rho_p(r^\prime) r^\prime dr^\prime \right].
\end{equation}
The centrifugal potential $V_L$ is written as
\begin{equation}
V_L = \frac{\hbar^2}{2m_\mu r^2} \left(\ell +\frac{1}{2}\right)^2 ,
\end{equation}
where $m_\mu$ is the reduced mass, and the Langer modification factor~\cite{Langer37} is adopted.

\subsection{Nucleon density profiles}

Since the $\alpha$ cluster potential of Eq.~\eqref{eq:V_N} requires the information
on the density profile of the daughter nucleus, we rely on microscopic models for nuclear structure. 
In the present work, we consider the Skyrme SLy4 (zero-range)~\cite{CBHMS98}
and the Gogny D1S (finite-range)~\cite{BGG91} models as non-relativistic approaches
and the relativistic mean-field interaction DD-ME2 model of Ref.~\cite{LNVR05} 
as a relativistic approach.

The Skyrme force model is constructed based on nucleon-nucleon interactions having
dependence on the relative momentum and density, which reads
\begin{eqnarray}
\label{eq:skyint}
v_{ij}^{} &=& t_0^{} \left(1 + x_0^{} P_\sigma \right) 
\delta \left(\mathbf{r}_i^{} - \mathbf{r}_j^{} \right) 
\nonumber \\ && \mbox{} 
+ \frac{t_1^{}}{2} \left(1+x_1^{} P_\sigma \right)  
\nonumber \\ && \mbox{} \quad \times
\left[
\delta \left(\mathbf{r}_i^{} - \mathbf{r}_j^{} \right)  \mathbf{k}^2 
+ \mathbf{k}^{\prime2} \delta \left(\mathbf{r}_i^{} - \mathbf{r}_j^{} \right) \right] 
\nonumber \\ && \mbox{} 
+ t_2^{} (1+ x_2^{} P_\sigma)\, \mathbf{k}^\prime \cdot \delta (\mathbf{r}_i^{} - \mathbf{r}_j^{} )
\mathbf{k} 
\nonumber \\ && \mbox{} 
+ \frac{t_3}{6} \left(1+x_3P_\mathbf{\sigma} \right)\rho^{\alpha}\delta(\mathbf{r}_i -\mathbf{r}_j) 
\nonumber \\ && \mbox{} 
+ i \, W_0\, \mathbf{k}^{\prime} \delta(\mathbf{r}_i -\mathbf{r}_j)\times \mathbf{k}
\cdot (\bm{\sigma}_i + \bm{\sigma}_j) ,
\end{eqnarray}
where $P_\sigma$ is the spin exchange operator, and $\bm{\sigma}_i$ are the Pauli 
spin matrices.
Here, $\mathbf{k}$ and $\mathbf{k}^\prime$ are the relative momenta of two nucleons 
before and after interaction, respectively, and
$W_0$ is the strength of the spin-orbit coupling.
There are many versions of the parameter set $(t_i,x_i,W_0)$ and, in the present work, 
we use the SLy4 model compiled in Ref.~\cite{CBHMS98}.

Compared with the Skyrme force model, the Gogny force assumes finite-range 
nucleon-nucleon interactions and zero-range multi-body forces, which leads to~\cite{DG80}
\begin{eqnarray}
v_{12}^{} &=& 
\sum_{j=1,2} \exp \left\{-\frac{ \left(\mathbf{r}_1^{} - \mathbf{r}_2^{} \right)^2}{\mu_j^2} \right\}
\nonumber \\ && \mbox{} \qquad \times
\left( W_j + B_j P_\sigma - H_j P_\tau - M_j P_\sigma P_\tau \right) 
\nonumber \\ && \mbox{}
+ t_0^{} \left(1 + x_0^{} P_\sigma \right) \rho^{\alpha} 
\left( \frac{\mathbf{r}_1^{} + \mathbf{r}_2^{}}{2} \right)
\delta \left( \mathbf{r}_1^{} - \mathbf{r}_2^{} \right) 
\nonumber \\ && \mbox{}
+ i W_{LS} \,\mathbf{k}^\prime \delta \left(\mathbf{r}_1^{} - \mathbf{r}_2^{} \right)
\times \mathbf{k} \cdot \left( \bm{\sigma}_1^{} + \bm{\sigma}_2^{} \right) ,
\end{eqnarray}
where $P_\tau$ is the isospin exchange operator.
We use the parameter values known as the D1S model in Ref.~\cite{BGG91}.

For nucleon density distribution, we also use a relativistic mean-field model 
of Refs.~\cite{LNVR05,NPVR14}, which gives a satisfactory description for  
the properties of finite nuclei. 
In this model, the relativistic Lagrangian density is given by
\begin{eqnarray}
\mathcal{L} &=& \bar{\psi} \left( i \slashed{\partial} -m \right) \psi
+ \frac{1}{2} \partial^\mu \sigma \partial_\mu \sigma - \frac{1}{2} m_\sigma \sigma^2 
- g_\sigma \bar{\psi} \sigma \psi 
\nonumber \\ && \mbox{}
 - \frac{1}{4} \Omega^{\mu\nu} \Omega_{\mu\nu} 
 + \frac{1}{2}m_\omega^2\omega^2 
- g_\omega \bar{\psi} \gamma^\mu \omega_\mu \psi 
\nonumber \\ && \mbox{}
- \frac{1}{4}\vec{R}^{\mu\nu} \cdot \vec{R}_{\mu\nu}
 + \frac{1}{2}m_\rho^2 \vec{\rho}^{\,2} 
  - g_\rho \bar{\psi} \gamma^\mu \vec{\rho}_\mu \cdot \vec{\tau} \psi 
\nonumber \\ && \mbox{}  
-\frac{1}{4} F^{\mu\nu} F_{\mu\nu}
  - e\bar{\psi} \gamma^\mu A_\mu \frac{(1-\tau_3^{})}{2}\psi\,,
\end{eqnarray}
where $\Omega^{\mu\nu}$, $\vec{R}^{\mu\nu}$, and $F^{\mu\nu}$ are 
the field strength tensors of the $\omega$ vector meson field $\omega_\mu$,
the isovector $\rho$ vector meson field $\vec{\rho}_\mu$, and the photon field 
$A_\mu$, respectively.
Note that the coupling constants of mesons to the nucleon are density-dependent so
as to reproduce the properties of nuclear matter and finite nuclei.
In the present work, we adopt the parameter set given as the DD-ME2 model in
Ref.~\cite{LNVR05}.

Within the Skyrme and Gogny force models, we solve Schr\"{o}dinger-like equations
to obtain the density profile of a nucleus. 
On the other hand, in the relativistic mean field model, we solve the Dirac equation to get the 
density profile for a given nucleus.
Once the density profile is known, one can find the $\alpha$ potential for each nucleus
and the $\alpha$ decay lifetime can be computed.
Since the $\alpha$ potential in Eq.~\eqref{eq:V_alpha} contains 6 parameters, we determine
these parameters to the experimental data for the alpha decays of even-even nuclei ($\ell=0$) 
as we have done in Ref.~\cite{SLHO15}. 
Table~\ref{tb:param} shows the parameters for the nuclear $\alpha$ potential determined
in this manner. 
The potential parameters for each model are found to have similar magnitudes except 
the case of $\gamma$, which is correlated to the value of $\epsilon$.
The $\gamma$ term is related with the multi-body force and we choose $\epsilon = \frac{1}{3}$ 
in the Gogny D1S model reflecting the original $\epsilon$ value in the Gogny $NN$ interaction.

\begin{table}[t]
\begin{tabular}{ccccl}
\hline
\hline
Parameter &  SLy4  &  D1S & DD-ME2 & ~~Unit \\
\hline
 $\alpha$ & $-1484.58$  &   $-1499.04$     & $-1524.24$    & ~~MeV fm$^{3}$ \\
 $\beta$  & $1355.57$   &   $1248.80$      & $1289.04$     & ~~MeV fm$^{5}$ \\ 
 $\gamma$ & $1005.48$   &   $242.28$      & $1137.21$     & ~~MeV fm$^{6+\epsilon}$ \\
 $\delta$ & $53.87$     &  $30.75$        &$-41.84$       & ~~MeV fm$^{5}$ \\
 $\eta$   & $-210.15$   &  $-178.12$      &$-184.09$      & ~~MeV fm$^{5}$ \\
 $\epsilon$ & $1/6$     &  $1/3$          & $1/6$         & \\
\hline
\end{tabular}
\caption{Parameters for $\alpha$ particle potential in Eq.~\eqref{eq:V_N}.}
\label{tb:param}
\end{table}

\section{Results}\label{sec:results}

Equipped with the $\alpha$ potential obtained in the previous section,
the $\alpha$-decay half-lives of heavy nuclei can be estimated in the standard
way by using the WKB approximation. 
The half-life of the nuclear $\alpha$ decay is related to the decay width $\Gamma$ by 
\begin{equation}
T_{1/2}^{} = \frac{\hbar\ln 2}{\Gamma} ,
\end{equation}
where the decay width is given by
\begin{equation}
\Gamma = \mathcal{PF} \frac{\hbar^2}{4m_\mu}\exp \left[-2\int_{r_2}^{r_3}\,dr k(r) \right].
\end{equation}
Here, $\mathcal{P}$ is the preformation factor which illustrates the probability of $\alpha$
particle in the parent nuclei, and $\mathcal{F}$ is the assaulting frequency of 
the trapped $\alpha$ particle between two turning points $r_1^{}$ and $r_2^{}$. 
In this calculation, we use $\mathcal{P} = 1$ and the explicit expression for $\mathcal{F}$
can be found, for example, in Ref.~\cite{SLHO15}.
The distance between $r_2^{}$ and $r_3^{}$, i.e., $\abs{r_2^{} - r_3^{}}$, represents the
penetration width of the barrier through which $\alpha$ particle passes.
$k(r)$ corresponds to the wave number of the $\alpha$ particle inside the potential barrier,
\begin{equation}
k(r) = \sqrt{\frac{2m_\mu}{\hbar^2} \abs{Q - V(r)} } 
\end{equation} 
with $m_\mu$ being the reduced mass of the system.

\begin{table*}[t]
\caption{Observed $\alpha$ decay half-lives of heavy nuclei and the results of the
present work. Unless specified, $\ell = 0$ is understood.} 
\label{tb:heavy}
\begin{tabular}{c|cccccccc}
\hline\hline
$(Z,A)$       &  $Q_{\alpha}^{\text{Expt}}$ (MeV) & $T_{1/2}^{\text{Expt}}$  
              & $T_{1/2}^{\text{SLy4}} ~[\ell]$   
              & $T_{1/2}^{\text{D1S}} ~[\ell]$
              & $T_{1/2}^{\text{DD-ME2}} ~[\ell]$ 
              & Reference  \\
\hline
~$(118, 294)$~ &  ~$11.81 \pm 0.06$~  & ~$0.89_{-0.31}^{+1.07}$ ms~ 
& ~$0.50^{+0.18}_{-0.13}$ ms~  %% SLy4
& ~$0.61^{+0.22}_{-0.16}$ ms~  %% Gogny 
& ~$0.43^{+0.15}_{-0.11}$ ms~   %% DD-ME
& \cite{OULA06}              \\
%%
%%
%% $^{+}_{-}$   &
%%
$(116,293)$ &  $10.67 \pm 0.06$    & $53_{-19}^{+62}$ ms 
&  ~$65^{+28}_{-20}$ ms %% SLy4
&  $78^{+33}_{-23}$ ms   
&  $54^{+24}_{-16}$ ms   %% DD-ME
& \cite{OULA04b}            \\
$(116,292)$ &  $10.80 \pm 0.07$           & $18_{-6}^{+16}$ ms 
&  ~$31^{+16}_{-10}$ ms~ %% SLy4
& $38^{+19}_{-13}$ ms  
& $26^{+13}_{-9}$ ms %% DD-ME
& \cite{OULA04b}   \\
$(116,291)$ &  $10.89 \pm 0.07$    & $18_{-6}^{+22}$ ms 
& $19^{+9}_{-6}$ ms     %% SLy4
& $23^{+11}_{-7}$ ms  
& $16^{+8}_{-5}$ ms %% DD-ME
& \cite{OULA06}  \\
$(116,290)$ &  $11.00 \pm 0.08$           & $7.1_{-1.7}^{+3.2}$ ms
& ~$10.6^{+6.1}_{-3.8}$ ms~  %% SLy4
& $12.5^{+7.2}_{-4.5}$ ms  
& $8.6^{+5.0}_{-3.1}$ ms %% DD-ME
& \cite{OULA06}  \\
$(115,288)$ &  $10.61 \pm 0.06$    & $87_{-30}^{+105}$ ms 
& $51^{+21}_{-15}$ ms    %% SLy4
& $57^{+25}_{-17}$ ms  %% Gogny D1s
& $42^{+19}_{-13}$ ms %% DD-ME
& \cite{OULA04,OUDL05}  \\
$(115,287)$ &  $10.74 \pm 0.09$    & $32_{-14}^{+155}$ ms 
&~$25^{+17}_{-10}$ ms~       %% SLy4
& $28^{+20}_{-12}$ ms   
& $21^{+15}_{-9}$ ms %% DD-ME  
& \cite{OULA04,OUDL05}  \\
$(114,289)$ &  $9.96 \pm 0.06$   & $2.7_{-0.7}^{+1.4}$ s 
& $1.3^{+0.6}_{-0.4}$ s    %% SLy4
& $1.5^{+0.7}_{-0.5}$ s  
& $1.0^{+0.5}_{-0.3}$ s %% DD-ME  
& \cite{OULA04b}          \\
$(114,288)$ &  $10.09 \pm 0.07$    & $0.8_{-0.18}^{+0.32}$ s
 & ~$0.56^{+0.31}_{-0.20}$ s~  %% SLy4
 & $0.65^{+0.37}_{-0.23}$ s  
 & $0.46^{+0.26}_{-0.16}$ s  %% DD-ME  
 & \cite{OULA04b}           \\
$(114,287)$ &  $10.16 \pm 0.06$   & $0.48_{-0.09}^{+0.16}$ s
& $0.37^{+0.17}_{-0.12}$ s %% SLy4
& $0.42^{+0.20}_{-0.13}$ s  
& $0.31^{+0.15}_{-0.10}$ s   %% DD-ME
& \cite{OULA06}         \\
$(114,286)$ &  $10.33 \pm 0.06$           & $0.13_{-0.02}^{+0.04}$ s
&  ~$0.14^{+0.06}_{-0.04}$ s~ %% SLy4
& $0.15^{+0.07}_{-0.05}$  s 
& $0.12^{+0.05}_{-0.04}$  s %% DD-ME  
& \cite{OULA06}    \\
$(113,284)$ &  $10.15 \pm 0.06$  & $0.48_{-0.17}^{+0.58}$ s
& $0.20^{+0.09}_{-0.06}$ s  %% SLy4
& $0.23^{+0.10}_{-0.07}$ s  %% Gogny
& $0.28^{+0.13}_{-0.09}$ s [$\ell = 2$]  %% DD-ME      
& \cite{OULA04,OUDL05}      \\
$(113,283)$ &  $10.26 \pm 0.09$  & $100_{-45}^{+490}$ ms 
&~$106^{+77}_{-45}$ ms~    %% SLy4
& $120^{+89}_{-51}$ ms  %% Gogny
& $94^{+70}_{-40}$ ms  %% DD-ME
& \cite{OULA04,OUDL05}          \\
$(113,282)$ &  $10.83 \pm 0.08$   & $73_{-29}^{+134}$ ms 
& $106^{+62}_{-38}$ ms [$\ell = 6$]%% SLy4   
& $121^{+73}_{-45}$ ms [$\ell = 6$]  %% Gogny
& $93^{+55}_{-34}$ ms  [$\ell = 6$] %% DD-ME
& \cite{OULA07}        \\
$(112,285)$ &  $9.29 \pm 0.06$            & $34_{-9}^{+17}$ s
& $27^{+14}_{-10}$ ms %% SLy4   
& $30^{+16}_{-10}$ s  %% Gogny missing
& $22^{+13}_{-8}$ s  %% DD-ME    
& \cite{OULA04b}       \\
$(112,283)$ &  $9.67 \pm 0.06$    & $3.8_{-0.7}^{+1.2}$ s 
& $2.0^{+1.0}_{-0.7}$ s  %% SLy4
& $2.3^{+1.2}_{-0.8}$ s   %% Gogny missing
& $1.8^{+0.9}_{-0.6}$ s   %% DD-ME
& \cite{OULA06}   \\
$(111,280)$ &  $9.87 \pm 0.06$            & $3.6_{-1.3}^{+4.3}$ s
& $1.4^{+0.7}_{-0.4}$ s [$\ell = 4$]%% SLy4
& $1.6^{+0.8}_{-0.5}$ s [$\ell = 4$]  %% Gogny
& $7.2^{+3.4}_{-2.3}$ s [$\ell = 6$]  %% DD-ME 
& \cite{OULA04,OUDL05}              \\
$(111,279)$ &  $10.52 \pm 0.16$  & $170_{-80}^{+810}$ ms 
& ~$157^{+251}_{-95}$ ms [$\ell = 6$]%% SLy4
& $176^{+276}_{-106}$ ms [$\ell = 6$]  
& $138^{+219}_{-83}$ ms  [$\ell = 6$]  %% DD-ME 
& \cite{OULA04,OUDL05}      \\
$(111,278)$ &  $10.89 \pm 0.08$   & $4.2_{-1.7}^{+7.5}$ ms 
& $3.5^{+1.9}_{-1.3}$ ms  [$\ell = 4$] %% SLy4
& $3.9^{+2.2}_{-1.4}$ ms  [$\ell = 4$]   %% Gogny
& $3.2^{+1.8}_{-1.1}$ ms  [$\ell = 4$]   %% DD-ME   
& \cite{OULA07}     \\ 
$(110,279)$ &  $9.84 \pm 0.06$            & $0.20_{-0.04}^{+0.05}$ s
& $0.15^{+0.07}_{-0.05}$ s   %% SLy4
& $0.17^{+0.08}_{-0.05}$ s   %% Gogny missing
& $0.13^{+0.06}_{-0.04}$ s   %% DD-ME 
& \cite{OULA06}         \\
$(109,276)$ &  $9.85 \pm 0.06$    & $0.72_{-0.25}^{+0.97}$ s
& $0.37^{+0.17}_{-0.12}$ s  [$\ell = 4$] %% SLy4
& $0.41^{+0.19}_{-0.13}$ s  [$\ell = 4$] %% Gogny
& $0.33^{+0.16}_{-0.10}$ s  [$\ell = 4$] %% DD-ME  
& \cite{OULA04,OUDL05}         \\
$(109,275)$ &  $10.48 \pm 0.09$  & $9.7_{-4.4}^{+46}$ ms 
& $8.7^{+5.9}_{-3.5}$ ms   [$\ell = 4$]%% SLy4
& $9.4^{+6.6}_{-3.8}$ ms   [$\ell = 4$]
& $7.9^{+5.4}_{-3.2}$ ms   [$\ell = 4$]%% DD-ME    
& \cite{OULA04,OUDL05}      \\
$(109,274)$ &  $9.95 \pm 0.10$  & $440_{-170}^{+810}$ ms 
& $220^{+195}_{-99}$ ms [$\ell = 4$]  %% SLy4
& $242^{+211}_{-112}$ ms[$\ell = 4$]   %% Gogny 
& $200^{+170}_{-94}$ ms [$\ell = 4$] %% DD-ME
&\cite{OULA07}  \\
$(108,275)$ &  $9.44 \pm 0.06$   & $0.19_{-0.07}^{+0.22}$ s
& $0.46^{+0.23}_{-0.15}$ s  %% SLy4
& $0.51^{+0.25}_{-0.17}$ s    %% Gogny
& $0.42^{+0.21}_{-0.14}$ s    %% DD-ME    
& \cite{OULA06}         \\
$(107,272)$ &  $9.15 \pm 0.06$            & $9.8_{-3.5}^{+11.7}$ s
& $9.0^{+4.7}_{-3.1}$ s  [$\ell = 4$]%% SLy4 
& $9.7^{+5.1}_{-3.3}$ s  [$\ell = 4$]   %% Gogny
& $7.9^{+4.1}_{-2.7}$ s  [$\ell = 4$] %% DD-ME
& \cite{OULA04,OUDL05}             \\
$(107,270)$ &  $9.11 \pm 0.08$      & $61_{-28}^{+292}$ s 
& $73^{+58}_{-30}$ s [$\ell = 6$] %% SLy4
& $84^{+64}_{-36}$ s [$\ell = 6$]  %% Gogny  
& $70^{+54}_{-30}$ s [$\ell = 6$]  %% DD-ME  
& \cite{OULA07}     \\
$(106,271)$ &  $8.67 \pm 0.08$  & $1.9_{-0.6}^{+2.4}$ min
& ~$2.10^{+1.77}_{-0.95}$ min [$\ell = 4$]  
& ~$2.27^{+1.99}_{-1.02}$ min [$\ell = 4$]  
& ~$1.83^{+1.54}_{-0.83}$ min [$\ell = 4$] %% DD-ME         
& \cite{OULA06}       \\
\hline
RMSD &  - &  - & $0.209$ & $0.198$ &  $0.218$ & \\
\hline\hline
\end{tabular}
\end{table*}

The heavy nuclei under study in the present work are neutron-rich but are located on the 
neutron-deficient side of beta-stability.
Thus, $\beta$ decay does not occur for these nuclei.
Table~\ref{tb:heavy} shows our results on the observed $\alpha$ decay half-lives of heavy nuclei.
Our results are obtained with the three models for nuclear density profiles 
and are compared with experimental data.
The theoretical uncertainties shown in the table come from those of the experimental $Q$
values.
The obtained half-lives depend on the relative orbital angular momentum $\ell$.
We assume $\ell=0$ for even-even decay cases
but we allow the variation of $\ell$ in other types of decay processes. 
The value of $\ell$ which minimizes the difference with the experimental data for the half-life 
is explicitly shown in Table~\ref{tb:heavy}. 
The results for half-lives without the value of $\ell$ are obtained with $\ell = 0$.
Compared with the previous results given in Ref.~\cite{SLHO15} which used a simple
Fermi density profile, using realistic proton distribution improves the rms deviation (RMSD) in 
$\alpha$ decay lifetimes as shown in the table, which is defined by
\begin{equation}
\mbox{RMSD} = \sqrt{\frac{1}{N-1}  \sum_i 
\left( \log_{10} \left[\frac{T_{i}^{\rm expt.}}{T_i^{\rm cal.}} \right] \right)^2 },
\end{equation}
where $N$ is the total number of data.
This indicates that the density profile of the neutron-rich heavy nuclei deviates
from the simple Fermi density profile and its effect
should be considered to get more realistic results.

Presented in Table~\ref{tb:pre} are our predictions on the half-lives of unobserved $\alpha$ decays 
of superheavy elements.
In this case, the $Q$ values are estimated by using the LDM and the local formula as described 
in Sec.~\ref{sec:qval}.
We assume $\ell=0$ for simplicity as there is no information on these processes.%
\footnote{If $\ell \neq 0$, the potential barrier width becomes larger than the case of $\ell=0$
and the lifetime becomes longer. 
For example, when $Q = 11 \sim 14$~MeV, if we use the Gogny D1S model, the enhancement 
factors for the half-life become $1.06$, $1.61$, $2.16$, $4.40$, and $8.08$ 
as we increase the value of $\ell$ from $1$ to $5$. Other models give similar results.}
Note that the half-lives from the D1S calculation are longer than the ones from SLy4 and 
DD-ME2 calculations. 
We found that this is mostly caused by the differences in parameters given in Table~\ref{tb:param}.

\begin{table*}[t]
\caption{Predictions on the $\alpha$ decay lifetimes for unobserved superheavy elements with $Q$ 
values from the LDM (case II) and from the local formula.}
\begin{tabular}{c|cccc|cccc}
\hline\hline
\multirow{2}{*}{~Nuclei $(Z,A)$~}  &
  ~$Q$ (MeV)~      & 
\multirow{2}{*}{$T_{1/2}^{\text{SLy4}}$ (s)}   & 
\multirow{2}{*}{~$T_{1/2}^{\text{D1S}}$ (s)~}  & 
\multirow{2}{*}{~$T_{1/2}^{\text{DD-ME2}}$ (s)~} & 
$Q$ (MeV)~ &
\multirow{2}{*}{$T_{1/2}^{\text{SLy4}}$ (s)} &
\multirow{2}{*}{~$T_{1/2}^{\text{D1S}}$ (s)~} &
\multirow{2}{*}{ ~$T_{1/2}^{\text{DD-ME2}}$ (s)~}
 \\
 & LDM &  &  &  & Local formula & & & \\
\hline
(122, 307) &  12.594
           & $ 9.467\times 10^{-5}$  %% SLy4
           & $ 9.982\times 10^{-5}$   %% D1S
           & $ 6.999\times 10^{-5}$  %% DD-ME
           & 12.289
           & $ 4.340\times 10^{-4}$  %% SLy4
           & $ 4.514\times 10^{-4}$   %% D1S
           & $ 3.194\times 10^{-4}$  %% DD-ME 
           \\
(122, 306) &  12.729
           & $ 5.649\times 10^{-5}$  %% SLy4
           & $ 5.836\times 10^{-5}$   %% D1S
           & $ 4.183\times 10^{-5}$  %% DD-ME
           & 12.420
           & $ 2.517\times 10^{-4}$  %% SLy4
           & $ 2.688\times 10^{-4}$   %% D1S
           & $ 1.891\times 10^{-4}$  %% DD-ME 
           \\
(122, 305) & 12.853
           & $ 3.334\times 10^{-5}$  %% SLy4
           & $ 3.607\times 10^{-5}$   %% D1S
           & $ 2.525\times 10^{-5}$  %% DD-ME
           & 12.550
           & $ 1.402\times 10^{-4}$  %% SLy4
           & $ 1.539\times 10^{-4}$   %% D1S
           & $1.073\times 10^{-4}$  %% DD-ME 
           \\
(122, 304) & 12.986
           & $ 1.931\times 10^{-5}$  %% SLy4
           & $ 2.100\times 10^{-5}$   %% D1S
           & $ 1.480\times 10^{-5}$  %% DD-ME
           & 12.679
           & $ 7.919\times 10^{-5}$  %% SLy4
           & $ 8.911\times 10^{-5}$  %% D1S
           & $ 6.193\times 10^{-5}$  %% DD-ME 
           \\
(122, 303) &  13.108
           & $ 1.145\times 10^{-5}$  %% SLy4
           & $ 1.300\times 10^{-5}$   %% D1S
           & $ 9.047\times 10^{-6}$  %% DD-ME
           & 12.807
           & $ 4.646\times 10^{-5}$  %% SLy4
           & $ 5.237\times 10^{-5}$   %% D1S
           & $3.593\times 10^{-5}$  %% DD-ME 
           \\
(122, 302) & 13.239
           & $ 6.692\times 10^{-6}$  %% SLy4
           & $ 7.539\times 10^{-6}$   %% D1S
           & $ 5.339\times 10^{-6}$  %% DD-ME
           & 12.935
           & $ 2.646\times 10^{-5}$  %% SLy4
           & $ 3.000\times 10^{-5}$   %% D1S
           & $2.099\times 10^{-5}$  %% DD-ME 
           \\
\hline
(121, 306) &  12.114
           & $ 5.360\times 10^{-4}$  %% SLy4
           & $ 5.522\times 10^{-4}$   %% D1S
           & $ 3.846\times 10^{-4}$  %% DD-ME
           & 11.853
           & $ 2.104\times 10^{-3}$  %% SLy4
           & $ 2.175\times 10^{-3}$   %% D1S
           & $1.509\times 10^{-3}$  %% DD-ME 
           \\
(121, 305) & 12.250
           & $ 2.948\times 10^{-4}$  %% SLy4
           & $ 3.093\times 10^{-4}$   %% D1S
           & $ 2.170\times 10^{-4}$  %% DD-ME
           & 11.985
           & $ 1.143\times 10^{-3}$  %% SLy4
           & $ 1.212\times 10^{-3}$   %% D1S
           & $8.467\times 10^{-4}$  %% DD-ME 
           \\
(121, 304) &  12.367
           & $ 1.664\times 10^{-4}$  %% SLy4
           & $ 1.831\times 10^{-4}$   %% D1S
           & $ 1.274\times 10^{-4}$  %% DD-ME
           & 12.117
           & $ 6.082\times 10^{-4}$  %% SLy4
           & $ 6.787\times 10^{-4}$   %% D1S
           & $4.700\times 10^{-4}$  %% DD-ME 
           \\
(121, 303) &  12.511
           & $ 9.077\times 10^{-5}$  %% SLy4
           & $ 1.030\times 10^{-4}$   %% D1S
           & $ 7.119\times 10^{-5}$  %% DD-ME
           & 12.248
           & $ 3.317\times 10^{-4}$  %% SLy4
           & $ 3.794\times 10^{-4}$   %% D1S
           & $2.593\times 10^{-4}$  %% DD-ME 
           \\
(121, 302) &  12.636
           & $ 5.323\times 10^{-5}$  %% SLy4
           & $ 6.026\times 10^{-5}$   %% D1S
           & $ 4.191\times 10^{-5}$  %% DD-ME
           & 12.378
           & $ 1.834\times 10^{-4}$  %% SLy4
           & $ 2.093\times 10^{-4}$   %% D1S
           & $1.439\times 10^{-4}$  %% DD-ME
            \\
(121, 301) &  12.769
           & $ 2.976\times 10^{-5}$  %% SLy4
           & $ 3.401\times 10^{-5}$   %% D1S
           & $ 2.378\times 10^{-5}$  %% DD-ME
           & 12.508
           & $ 1.027\times 10^{-4}$  %% SLy4
           & $ 1.169\times 10^{-4}$   %% D1S
           & $8.201\times 10^{-5}$  %% DD-ME
            \\
\hline
(120, 304) & 11.790
           & $ 1.567\times 10^{-3}$  %% SLy4
           & $ 1.650\times 10^{-3}$   %% D1S
           & $ 1.167\times 10^{-3}$  %% DD-ME
           & 11.546
           & $ 5.792\times 10^{-3}$  %% SLy4
           & $ 6.146\times 10^{-3}$   %% D1S
           & $4.349\times 10^{-3}$  %% DD-ME 
           \\
(120, 303) &  11.918
           & $ 8.584\times 10^{-4}$  %% SLy4
           & $ 9.358\times 10^{-4}$   %% D1S
           & $ 6.494\times 10^{-4}$  %% DD-ME
           & 11.679
           & $ 2.987\times 10^{-3}$  %% SLy4
           & $ 3.331\times 10^{-3}$   %% D1S
           & $2.289\times 10^{-3}$  %% DD-ME 
           \\
(120, 302) &  12.055
           & $ 4.456\times 10^{-4}$  %% SLy4
           & $ 5.025\times 10^{-4}$   %% D1S
           & $ 3.459\times 10^{-4}$  %% DD-ME
           & 11.812
           & $ 1.561\times 10^{-3}$  %% SLy4
           & $ 1.761\times 10^{-3}$   %% D1S
           & $1.217\times 10^{-3}$  %% DD-ME 
           \\
(120, 301) & 12.181
           & $ 2.491\times 10^{-4}$  %% SLy4
           & $ 2.816\times 10^{-4}$   %% D1S
           & $ 1.959\times 10^{-4}$  %% DD-ME
           & 11.944
           & $ 8.288\times 10^{-4}$  %% SLy4
           & $ 9.395\times 10^{-4}$   %% D1S
           & $ 6.575\times 10^{-4}$  %% DD-ME 
           \\
(120, 300) & 12.317
           & $ 1.342\times 10^{-4}$  %% SLy4
           & $ 1.523\times 10^{-4}$   %% D1S
           & $ 1.068\times 10^{-4}$  %% DD-ME
           & 12.076
           & $ 4.465\times 10^{-4}$  %% SLy4
           & $ 5.053\times 10^{-4}$   %% D1S
           & $ 3.520\times 10^{-4}$  %% DD-ME 
           \\
(120, 299) & 12.442
           & $ 7.735\times 10^{-5}$  %% SLy4
           & $ 8.978\times 10^{-5}$   %% D1S
           & $ 6.175\times 10^{-5}$  %% DD-ME
           & 12.207
           & $ 2.436\times 10^{-4}$  %% SLy4
           & $ 2.817\times 10^{-4}$   %% D1S
           & $ 1.957\times 10^{-4}$  %% DD-ME 
           \\
\hline
(119, 298) &  11.973
           & $ 4.022\times 10^{-4}$  %% SLy4
           & $ 4.688\times 10^{-4}$   %% D1S
           & $ 3.243\times 10^{-4}$  %% DD-ME
           & 11.772
           & $ 1.131\times 10^{-3}$  %% SLy4
           & $ 1.322\times 10^{-3}$   %% D1S
           & $ 8.986\times 10^{-4}$  %% DD-ME 
           \\
(119, 297) &  12.109
           & $ 2.119\times 10^{-4}$  %% SLy4
           & $ 2.415\times 10^{-4}$   %% D1S
           & $ 1.706\times 10^{-4}$  %% DD-ME
           & 11.904
           & $ 5.932\times 10^{-4}$  %% SLy4
           & $ 1.610\times 10^{-3}$   %% D1S
           & $ 4.795\times 10^{-4}$  %% DD-ME 
           \\
(119, 296)  &  12.234
           & $ 1.181\times 10^{-4}$  %% SLy4
           & $ 1.340\times 10^{-4}$   %% D1S
           & $ 9.719\times 10^{-5}$  %% DD-ME
           & 12.036
           & $ 3.147\times 10^{-4}$  %% SLy4
           & $ 3.587\times 10^{-4}$   %% D1S
           & $ 2.593\times 10^{-4}$  %% DD-ME 
           \\
(119, 295)  &  12.368
           & $ 6.172\times 10^{-5}$  %% SLy4
           & $ 7.814\times 10^{-5}$   %% D1S
           & $ 5.316\times 10^{-5}$  %% DD-ME
           & 12.167
           & $ 1.643\times 10^{-4}$  %% SLy4
           & $ 1.913\times 10^{-4}$   %% D1S
           & $ 1.405\times 10^{-4}$  %% DD-ME 
           \\
(119, 294)  &  12.492
           & $ 3.425\times 10^{-5}$  %% SLy4
           & $ 4.112\times 10^{-5}$   %% D1S
           & $ 2.983\times 10^{-5}$  %% DD-ME
           & 12.297
           & $ 8.668\times 10^{-5}$  %% SLy4
           & $ 1.044\times 10^{-4}$   %% D1S
           & $ 7.549\times 10^{-5}$  %% DD-ME 
           \\
(119, 293) & 12.625
           & $ 1.874\times 10^{-5}$  %% SLy4
           & $ 2.264\times 10^{-5}$   %% D1S
           & $ 1.646\times 10^{-5}$  %% DD-ME
           & 12.427
           & $ 4.775\times 10^{-5}$  %% SLy4
           & $ 5.767\times 10^{-5}$   %% D1S
           & $ 4.168\times 10^{-5}$  %% DD-ME 
           \\
\hline
(118, 298) & 11.393 
           & $ 4.077\times 10^{-3}$  %% SLy4
           & $ 4.600\times 10^{-3}$   %% D1S
           & $ 3.215\times 10^{-3}$  %% DD-ME
           & 11.197
           & $ 1.206\times 10^{-2}$  %% SLy4
           & $ 1.373\times 10^{-2}$   %% D1S
           & $ 9.535\times 10^{-3}$  %% DD-ME 
           \\
(118, 297) & 11.522
           & $ 2.126\times 10^{-3}$  %% SLy4
           & $ 2.488\times 10^{-3}$   %% D1S
           & $ 1.699\times 10^{-3}$  %% DD-ME
           & 11.332
           & $ 5.977\times 10^{-3}$  %% SLy4
           & $ 7.008\times 10^{-3}$   %% D1S
           & $ 4.774\times 10^{-3}$  %% DD-ME 
           \\
(118, 296) &  11.660
           & $ 1.068\times 10^{-3}$  %% SLy4
           & $ 1.238\times 10^{-3}$   %% D1S
           & $ 8.599\times 10^{-4}$  %% DD-ME
           & 11.466
           & $ 3.013\times 10^{-3}$  %% SLy4
           & $ 3.481\times 10^{-3}$   %% D1S
           & $ 2.423\times 10^{-3}$  %% DD-ME 
           \\
(118, 295) & 11.787
           & $ 5.640\times 10^{-4}$  %% SLy4
           & $ 6.577\times 10^{-4}$   %% D1S
           & $ 4.692\times 10^{-4}$  %% DD-ME
           & 11.600
           & $ 1.500\times 10^{-3}$  %% SLy4
           & $ 1.762\times 10^{-3}$   %% D1S
           & $ 1.244\times 10^{-3}$  %% DD-ME 
           \\
(118, 294) &  11.924
           & $ 2.824\times 10^{-4}$  %% SLy4
           & $ 8.069\times 10^{-4}$   %% D1S
           & $ 2.412\times 10^{-4}$  %% DD-ME
           & 11.733
           & $ 7.515\times 10^{-4}$  %% SLy4
           & $ 9.050\times 10^{-4}$   %% D1S
           & $ 6.387\times 10^{-4}$  %% DD-ME 
           \\
(118, 293) &  12.050
           & $ 1.516\times 10^{-4}$  %% SLy4
           & $ 1.835\times 10^{-4}$   %% D1S
           & $ 1.305\times 10^{-4}$  %% DD-ME
           & 11.865
           & $ 3.832\times 10^{-4}$  %% SLy4
           & $ 4.644\times 10^{-4}$   %% D1S
           & $ 3.289\times 10^{-4}$  %% DD-ME 
           \\
\hline
(117, 298) &  10.779
           & $ 6.202\times 10^{-2}$  %% SLy4
           & $ 7.032\times 10^{-2}$   %% D1S
           & $ 4.795\times 10^{-2}$  %% DD-ME
           & 10.920
           & $ 1.678\times 10^{-1}$  %% SLy4
           & $ 1.916\times 10^{-1}$   %% D1S %% missing
           & $ 1.311\times 10^{-1}$  %% DD-ME 
           \\
(117, 297) &  10.920
           & $ 2.837\times 10^{-2}$  %% SLy4
           & $ 3.274\times 10^{-2}$   %% D1S
           & $ 2.236\times 10^{-2}$  %% DD-ME
           & 10.749
           & $ 7.769\times 10^{-2}$  %% SLy4
           & $ 9.001\times 10^{-2}$   %% D1S
           & $ 6.129\times 10^{-2}$  %% DD-ME 
           \\
(117, 296) &  11.051
           & $ 1.409\times 10^{-2}$  %% SLy4
           & $ 1.666\times 10^{-2}$   %% D1S
           & $ 1.126\times 10^{-2}$  %% DD-ME
           & 10.886
           & $ 3.620\times 10^{-2}$  %% SLy4
           & $ 4.330\times 10^{-2}$   %% D1S
           & $ 2.903\times 10^{-2}$  %% DD-ME 
           \\
(117, 295) &  11.192
           & $ 6.660\times 10^{-3}$  %% SLy4
           & $ 7.806\times 10^{-3}$   %% D1S
           & $ 5.400\times 10^{-3}$  %% DD-ME
           & 11.023
           & $ 1.735\times 10^{-2}$  %% SLy4
           & $ 2.035\times 10^{-2}$   %% D1S
           & $ 1.396\times 10^{-2}$  %% DD-ME 
           \\
(117, 294) & 11.321
           & $ 3.310\times 10^{-3}$  %% SLy4
           & $ 3.965\times 10^{-3}$   %% D1S
           & $ 6.634\times 10^{-3}$  %% DD-ME
           & 11.158
           & $ 8.146\times 10^{-3}$  %% SLy4
           & $ 9.736\times 10^{-3}$   %% D1S
           & $ 6.779\times 10^{-3}$  %% DD-ME 
           \\
(117, 293) &  11.460
           & $ 1.584\times 10^{-3}$  %% SLy4
           & $ 1.941\times 10^{-3}$   %% D1S
           & $ 1.325\times 10^{-3}$  %% DD-ME
           & 11.293
           & $ 3.885\times 10^{-3}$  %% SLy4
           & $ 4.752\times 10^{-3}$   %% D1S
           & $ 3.244\times 10^{-3}$  %% DD-ME 
           \\
\hline\hline
\end{tabular}\label{tb:pre}
\end{table*}

\begin{figure*}[t]
\includegraphics[scale=0.4]{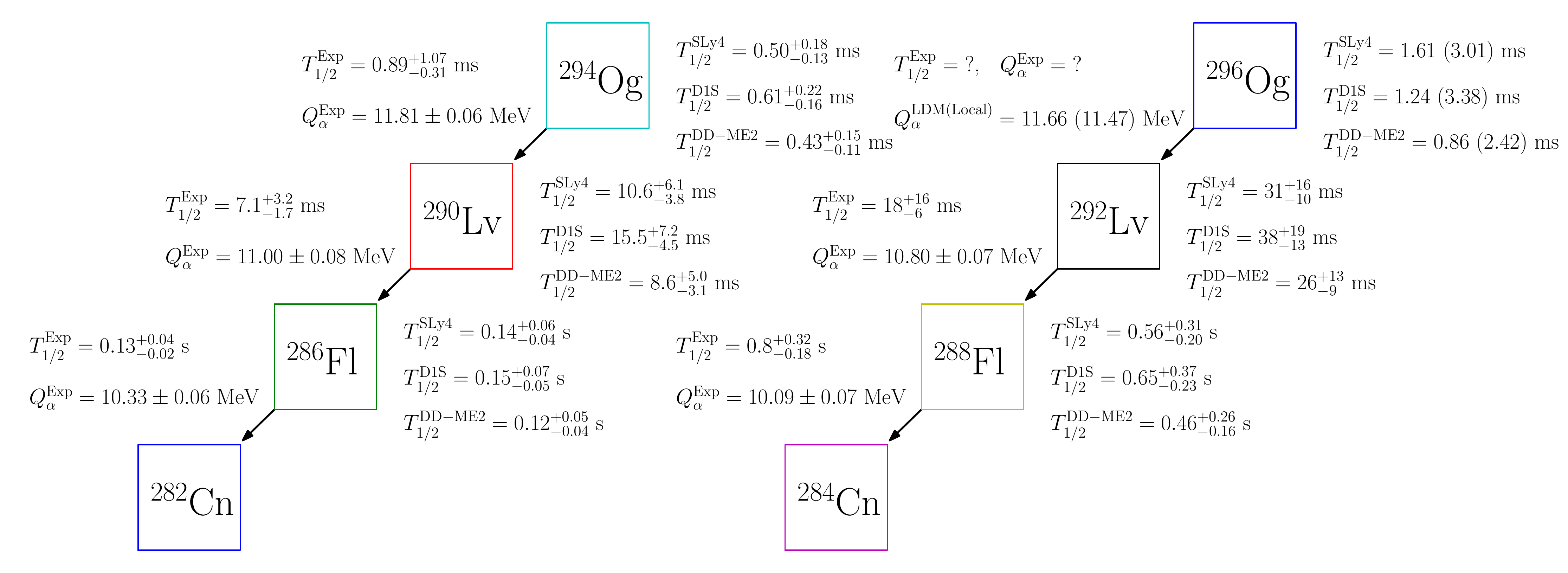}
\caption{
Float charts for $\alpha$ decay chains for \nuclide[294][118]{Og} and \nuclide[296][118]{Og}.
The measured half-life of \nuclide[286]{Fl} is about 0.13~s. Since the branching ratio of its $\alpha$
decay is about 60\%~\cite{OU15,OU15b}, however, the half-life of its $\alpha$ decay is about 0.22~s.}
\label{fig:chain}
\end{figure*}

Figure~\ref{fig:chain} shows one of the most important $\alpha$ decay chains of superheavy nuclei,
namely, the decay chains of \nuclide[294][118]{Og} and \nuclide[296][118]{Og}.
Our results successfully explain the $\alpha$ decay lifetimes in these two decay channels
compared with experimental results. 
The $\alpha$ decay of \nuclide[296][118]{Og} is yet to be discovered and the half-lives
for this decay given in Fig.~\ref{fig:chain} are our predictions. 
It should be noticed that the half-lives shown in Fig.~\ref{fig:chain} are calculated from
the nuclear $\alpha$ decay but the actual half-lives should be determined through the
competition with the spontaneous fission process. 
For example, in the case of \nuclide[286]{Fl}, although the measured half-life is 
$T^{\rm Exp.} \approx 0.13$~s, the branching ratio of the $\alpha$-decay is about 
60\%~\cite{OU15,OU15b}, which makes the $\alpha$-decay half-life close to 0.22~s.

\begin{figure}
\includegraphics[scale=0.45]{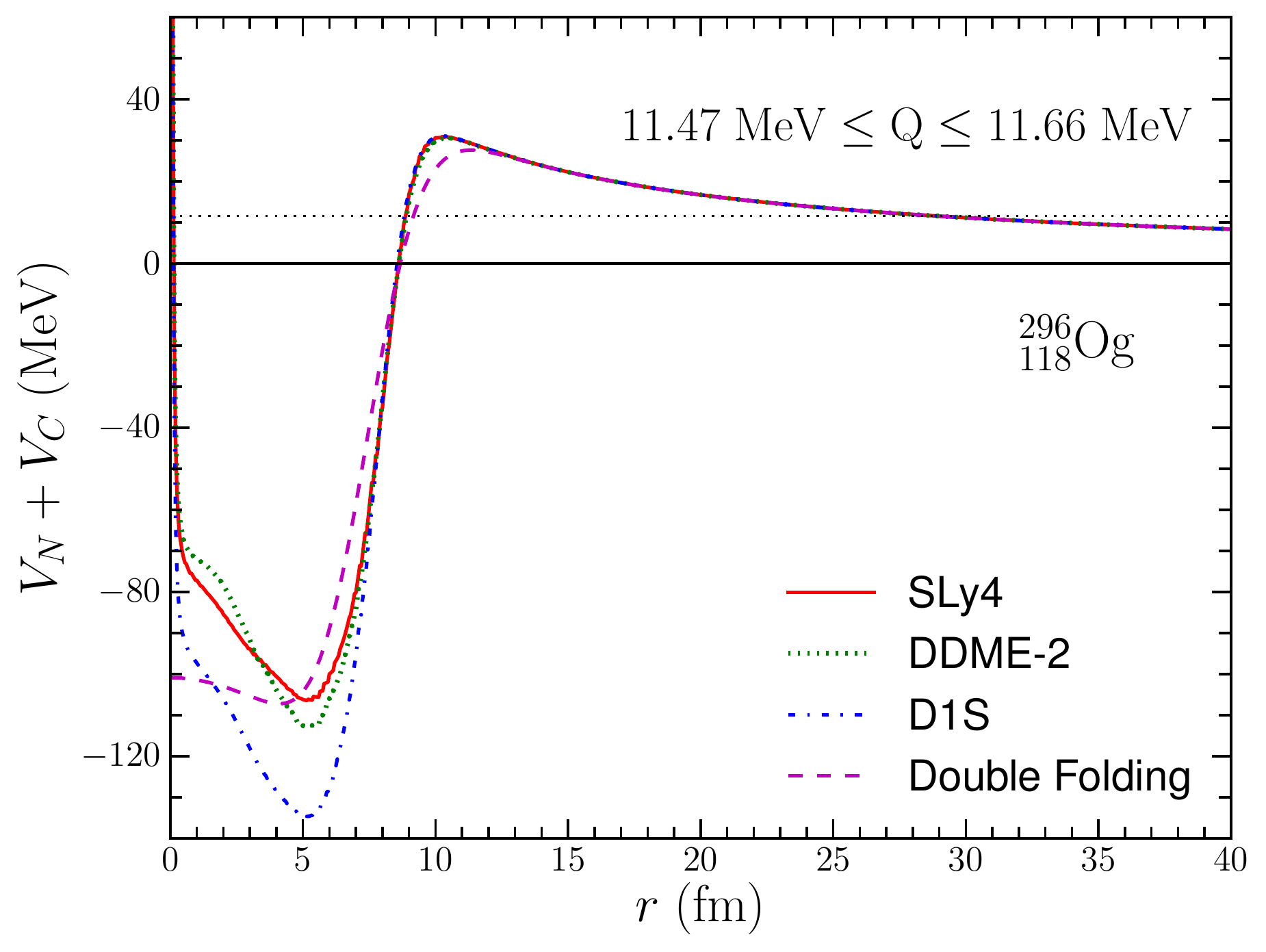}
\caption{
The $\alpha$ nuclear and Coulomb potentials, $V_N + V_C$, for \nuclide[296][118]{Og} in the models 
of the present work.  
The double folding potential for \nuclide[296][118]{Og} of Ref.~\cite{Mohr16} is also presented
for comparison. }
\label{fig:pot}
\end{figure}

Figure~\ref{fig:pot} shows the $\alpha$ potentials, $V_N + V_C$, used to calculate the half-life of 
\nuclide[296][118]{Og} in this work. 
The dotted line indicates the $Q$-values obtained in this work. 
The double folding potential is presented by the dashed line for comparison~\cite{Mohr16}. 
This shows that, although the details of the potentials in each model are quite different inside the
nucleus, the barrier widths corresponding to the obtained $Q$ values are relatively close to each other. 
The sightly lower barrier in Ref.~\cite{Mohr16} is compensated by a preformation factor of 0.09, 
finally leading to half-lives close to each other.

\section{Summary and Conclusion}\label{sec:conclusion}

In this paper, we have investigated the nuclear $\alpha$ decays of heavy nuclei based on
nuclear energy density functional.
We use a Skyrme-type force model to get the nuclear potential of the $\alpha$ particle inside
a nucleus as a functional of proton and neutron density profiles of the daughter nucleus.
These nucleon density profiles are obtained from the Skyrme SLy4, Gogny D1S, and
relativistic mean-field DD-ME2 models.
The parameters of the nuclear potential of the $\alpha$ are fitted for each density profile model
to measured $\alpha$ decay half-lives of heavy nuclei.
The results show that this approach improves the previous results reported in Ref.~\cite{SLHO15},
by reducing the RMS deviation from 0.238 to $0.198 \sim 0.218$.
In particular, we found that the Gogny D1S gives a better description among the models
considered in the present work.

Once all the parameters are fixed, we apply the model to predict half-lives of unobserved
$\alpha$ decays to get the estimations shown in Table~\ref{tb:pre}.
One interesting quantity is the half-life of \nuclide[296][118]{Og} as there are attempts
to synthesize this nuclide~\cite{Sobiczewski16}.
Our predictions on this decay are also shown in Fig.~\ref{fig:chain},
which shows our estimation of the $Q$ value as $Q^{\text{LDM}}=11.66$~MeV and
$Q^{\text{Local}} =11.47$~MeV.
Our predictions on the half-life of the $\alpha$ decay of this nuclide is in the range of
$0.86~\mbox{ms} \sim 3.48~\mbox{ms}$, which is in good agreement with the predictions of 
Ref.~\cite{Sobiczewski16} that gives $0.5~\mbox{ms} \sim 4.8~\mbox{ms}$ based on realistic
mass formulas and with the prediction of Ref.~\cite{Mohr16} which obtained 0.825~ms using the 
double-folding potential model. (See also Refs.~\cite{SPN16,Manjunatha16}.)

In the present work, we assumed that the potential for the $\alpha$ is isotropic.
However, in the case of heavy nuclei, the deformation effects should be included,
in particular, to understand its fine structure~\cite{DIL92,NR10}.
Therefore, improving the present model by including deformation and other microscopic
effects would be desired for a better understanding of nuclear $\alpha$ decays of superheavy
nuclei.

\acknowledgments

We are grateful to P. Papakonstantinou for providing us with density profiles
of nuclei obtained in the Gogny force model.
We also thank P. Mohr for providing his double folding potential for $\alpha$ decay and
many suggestions for this work.
The work of Y.O. was supported by Kyungpook National University Bokhyeon 
Research Fund, 2015.

\end{document}